\documentclass{appolb}
\usepackage{epsfig}
\usepackage{amsmath}
\usepackage{subfigure}

\begin{document}

\title{Light tetraquark state at nonzero temperature\thanks{%
Presented by A. Heinz at the Excited QCD Workshop, 31.1.-6.2.2010, in
Tatranska Lomnica (Slovakia)}}
\author{Achim Heinz$^{a}$, Stefan Str\"{u}ber$^{a}$, Francesco Giacosa$^{a}
$ and Dirk H. Rischke$^{a,b}$ 
\address{$^a$Institute for Theoretical Physics, Johann Wolfgang Goethe University,
Max-von-Laue-Str.\ 1, D--60438 Frankfurt am Main, Germany \and $^b$Frankfurt Institute for Advanced
Studies, Johann Wolfgang Goethe University,
Ruth-Moufang-Str.\ 1, D--60438 Frankfurt am Main, Germany} }
\maketitle

\begin{abstract}
We study the implications of a light tetraquark on the chiral phase
transition at nonzero temperature $T$: The behavior of the chiral and
four-quark condensates and the meson masses are studied in the scenario in
which the resonance $f_{0}(600)$ is described as a predominantly tetraquark
state. It is shown that the critical temperature is lowered and the
transition softened. Interesting mixing effects between tetraquark, and
quarkonium configurations take place.
\end{abstract}


%


\section{Introduction}

In the last decades theoretical and experimental work on light scalar
mesons with masses below $\sim 1.8\text{ GeV }$\cite{pdg} initiated an
intense debate about their nature. Quarkonia, tetraquark and mesonic
molecular assignments, together with the inclusion of a scalar glueball
state around $1.5\text{ GeV}$ as suggested by lattice simulations, have been
proposed and investigated in a variety of combinations and mixing patterns 
\cite{amslerrev}.

Nowadays evidence toward a full nonet of scalars below $1\text{ GeV}$ is
mounting: $f_{0}(600)$, $f_{0}(980)$, $a_{0}(980)$, and $K_{0}^{\ast }(800)$%
. An elegant way to explain such resonances is the tetraquark assignment
proposed by Jaffe long ago \cite{jaffe}. The reversed mass ordering is
naturally explained in this way and also decays can be successfully
reproduced \cite{maiani}. Within this context the lightest scalar resonance $%
f_{0}(600)$ is interpreted as a predominantly tetraquark state $1/2[u,d][%
\bar{u},\bar{d}],$ where the commutator indicates an antisymmetric flavour
configuration of the diquark.

The lightest quark-antiquark state, i.e., the chiral partner of the pion with
flavor wavefunction $\bar{n}n=\sqrt{1/2}(\bar{u}u+\bar{d}d)$, is then
predominantly identified with the broad resonance $f_{0}(1370)$. The fact
that scalar quarkonia are $p$-wave states supports this choice. According to
this picture quarkonia states, together with the scalar glueball, lie above 
$1$ GeV, see Ref.\ \cite{refs} and refs.\ therein.

It is natural to ask how the scenario outlined here affects the physics at
nonzero temperature $T$. It is in fact different from the usual assumptions
made in hadronic models at $T>0,$ where the chiral partner of the pion has a
mass of about $600$ MeV. Moreover, besides the chiral condensate, new
quantities emerge: a tetraquark condensate and the mixing of tetraquark and
quarkonium states in the vacuum and at nonzero $T.$ Remarkably, the mixing
angle increases for increasing $T$ and the behavior of the chiral condensate
is affected by the presence of the tetraquark field. Details can be found in
Ref.\ \cite{tqft}, on which these proceedings are based.

\section{The Model}

\label{2} We work with a simple chiral model with the following fields: the
pion triplet $\vec{\pi}$, the bare quarkonium field $\varphi \equiv \bar{n}n$%
, and bare tetraquark field $\chi \equiv \frac{1}{2}[u,d][\bar{u},\bar{d}]$.
The chiral potential was derived in Ref.\ \cite{tqmix}: 
\begin{equation}
V=\frac{\lambda }{4}\left( \varphi ^{2}+\vec{\pi}^{2}-F^{2}\right)
^{2}-\varepsilon \varphi +\frac{1}{2}M_{\chi }^{2}\chi ^{2}-g\chi (\varphi
^{2}+\vec{\pi}^{2})\text{ },  \label{pot}
\end{equation}%
where, besides the usual Mexican hat, the parameter $g$ describes the
interaction strength between the quark-antiquark fields and the tetraquark
field $\chi .$ In the limit $g\rightarrow 0$ the field $\chi $ decouples,
and a simple linear sigma model for $\varphi $ and $\vec{\pi}$ emerges. The
minimum of the potential (\ref{pot}) is, to order $O(\varepsilon )$:

\begin{equation}
\varphi _{0}\simeq \frac{F}{\sqrt{1-2g^{2}/(\lambda M_{\chi }^{2})}}+\frac{%
\varepsilon }{2\lambda F^{2}}\text{ },~~\chi _{0}=\frac{g}{M_{\chi }^{2}}%
\varphi _{0}^{2}\text{ },  \label{vev}
\end{equation}%
and $\vec{\pi}_{0}=0$. The condensate $\varphi _{0}$ is identified with the
pion decay constant $f_{\pi }=92.4\text{ MeV}$. Note that the tetraquark
condensate $\chi _{0}$ is proportional to $\varphi _{0}^{2}$: it is induced
by spontaneous symmetry breaking in the quarkonium sector. Shifting the
fields by their vacuum expectation values (vev's) $\varphi \rightarrow
\varphi +\varphi _{0}$ and $\chi \rightarrow \chi +\chi _{0}$, and expanding
around the minimum, we obtain up to second order in the fields 
\begin{equation}
V=\frac{1}{2}(\chi ,\varphi )\left( 
\begin{array}{cc}
M_{\chi }^{2} & -2g\varphi _{0} \\ 
-2g\varphi _{0} & M_{\varphi }^{2}%
\end{array}%
\right) \left( 
\begin{array}{c}
\chi  \\ 
\varphi 
\end{array}%
\right) +\frac{1}{2}M_{\pi }^{2}\vec{\pi}^{2}+\ldots \,,  \label{vexp}
\end{equation}%
where  
\begin{equation}
M_{\varphi }^{2}=\varphi _{0}^{2}\left( 3\lambda -\frac{2g^{2}}{M_{\chi }^{2}%
}\right) -\lambda F^{2},~~M_{\pi }^{2}=\frac{\epsilon }{\varphi _{0}} \text{ }.
\end{equation}%
Since the mass matrix has off-diagonal terms, the fields $\varphi $ and $\chi 
$ are not mass eigenstates. The mass eigenstates $H$ and $S$,
identified with the resonances $f_{0}(600)$ and $f_{0}(1370)$, respectively,
are obtained from an $SO(2)$ rotation of the fields $\varphi $ and $\chi $, 
\begin{equation}
\left( 
\begin{array}{c}
H \\ 
S%
\end{array}%
\right) =\left( 
\begin{array}{cc}
\cos \theta _{0} & \sin \theta _{0} \\ 
-\sin \theta _{0} & \cos \theta _{0}%
\end{array}%
\right) \left( 
\begin{array}{c}
\chi  \\ 
\varphi 
\end{array}%
\right) \,,\text{ }\theta _{0}=\frac{1}{2}\arctan \frac{4g\varphi _{0}}{%
M_{\varphi }^{2}-M_{\chi }^{2}} \text{ }. \label{diag2}
\end{equation}%
The tree-level masses of $H$ and $S$ are 
\begin{align}
M_{H}^{2}& =M_{\chi }^{2}\cos ^{2}\theta _{0}+M_{\varphi }^{2}\sin
^{2}\theta _{0}-2g\varphi _{0}\sin (2\theta _{0}), \\
M_{S}^{2}& =M_{\varphi }^{2}\cos ^{2}\theta _{0}+M_{\chi }^{2}\sin
^{2}\theta _{0}+2g\varphi _{0}\sin (2\theta _{0}).
\end{align}%
For the reasons discussed in the Introduction, the bare tetraquark is chosen
to be lighter than the bare quarkonium, thus: $M_{S}>M_{\varphi }>M_{\chi
}>M_{H}$. The state $H\equiv f_{0}(600)$ is the predominantly tetraquark
state, and the state $S\equiv f_{0}(1370)$ is the predominantly quarkonium
state. 

\section{Results and discussions}

\label{3} In order to investigate the nonzero $T$ behavior, we employ the
CJT formalism in the Hartree-Fock approximation \cite{cjt}; for
specification of the method in the case of mixing we refer to Ref.\ \cite{juergen}%
. The CJT-formalism leads to temperature-dependent masses $M_{S}(T)$, $%
M_{H}(T)$, $M_{\pi }(T)$, and a temperature-dependent mixing angle $\theta (T)
$. Moreover, both scalar-isoscalar fields have a $T$-dependent vev, for the
quarkonium $\varphi _{0}\rightarrow \varphi (T)$ and for the tetraquark $%
\chi _{0}\rightarrow \chi (T)$. For both fields the zero-temperature limits are $%
\varphi (0)=\varphi _{0}$ and $\chi (0)=\chi _{0}$.

When the tetraquark decouples (limit $g\rightarrow 0$), $S$ is a pure
quarkonium and $H$ is a pure tetraquark state. The transition is crossover for 
$M_{S}\leq 0.95\text{ GeV}$ and first order above this value. This is a well-established 
result, see e.g. Ref.\ \cite{stefan}. The fact that a heavy chiral
partner (i.e., mass larger $1\text{ GeV}$) of the pion leads to a first
order phase transition disagrees with lattice QCD calculations \cite%
{karschfodor}.

The inclusion of the tetraquark state changes this conclusion as shown in
Fig.\ 1: In Fig.\ 1.a $M_{H}=0.4\text{ GeV}$ is fixed and the parameters $M_{S}
$ and $g$ are varied. In Fig.\ 1.b the behavior of the quark condensate for
fixed $M_{S}=1.0\text{ GeV}$ and $M_{H}=0.4\text{ GeV}$ is shown for
different values of the parameter $g$. One observes that for increasing
values of the coupling $g$ the critical temperature $T_{c}$ decreases: while $%
T_{c}=250\text{ MeV}$ for $g\rightarrow 0$, a value $T_{c}\simeq 200\text{
MeV}$ is obtained for $g=2.0\text{ GeV}$. Also the order of the phase transition
is affected: when increasing the parameter $g$, the first-order transition is
softened and, if the coupling is large enough, becomes a crossover.

\begin{figure}[tbp]
\begin{center}
\includegraphics[scale=0.4965] {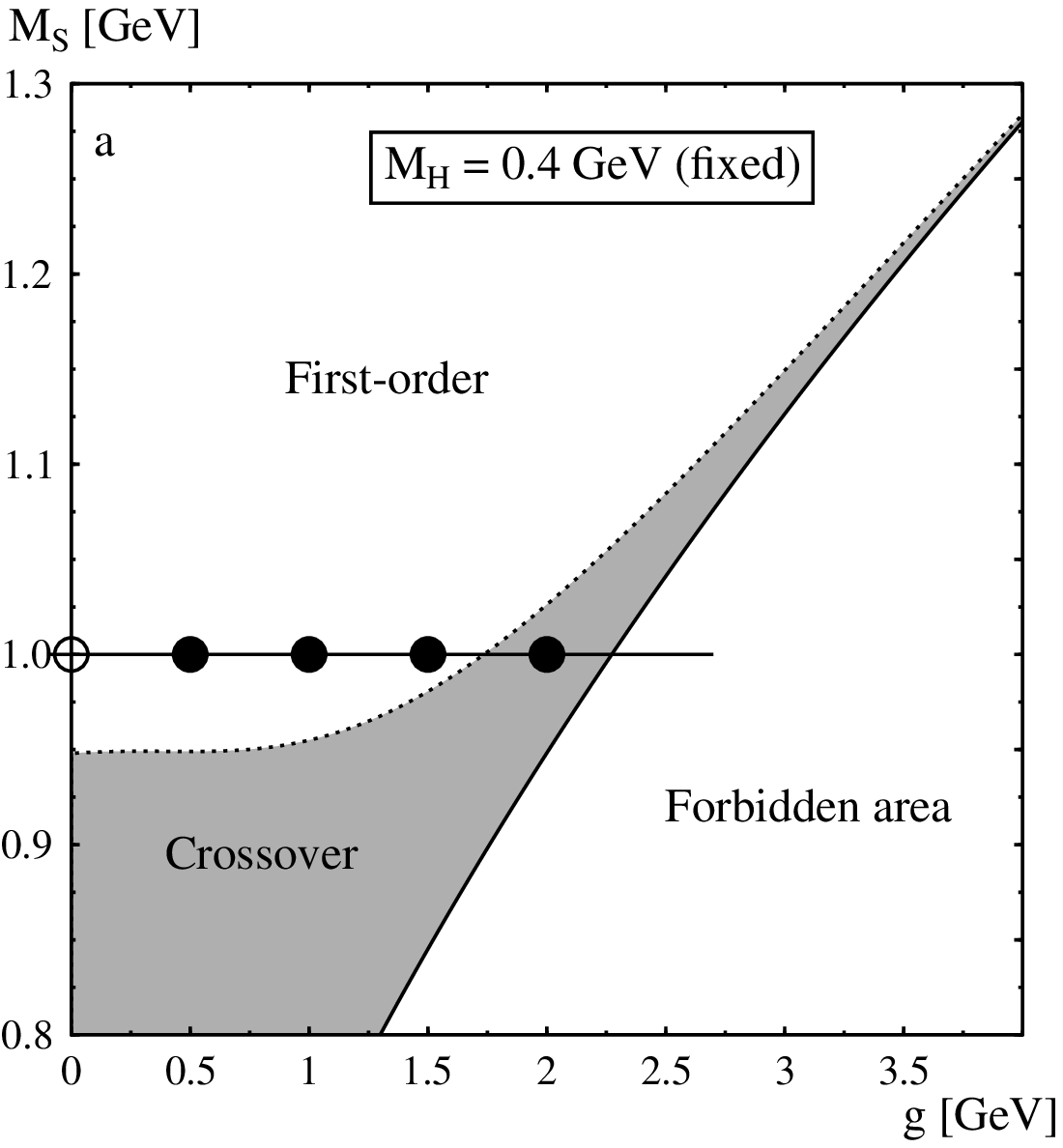} %
\includegraphics[scale=0.4965] {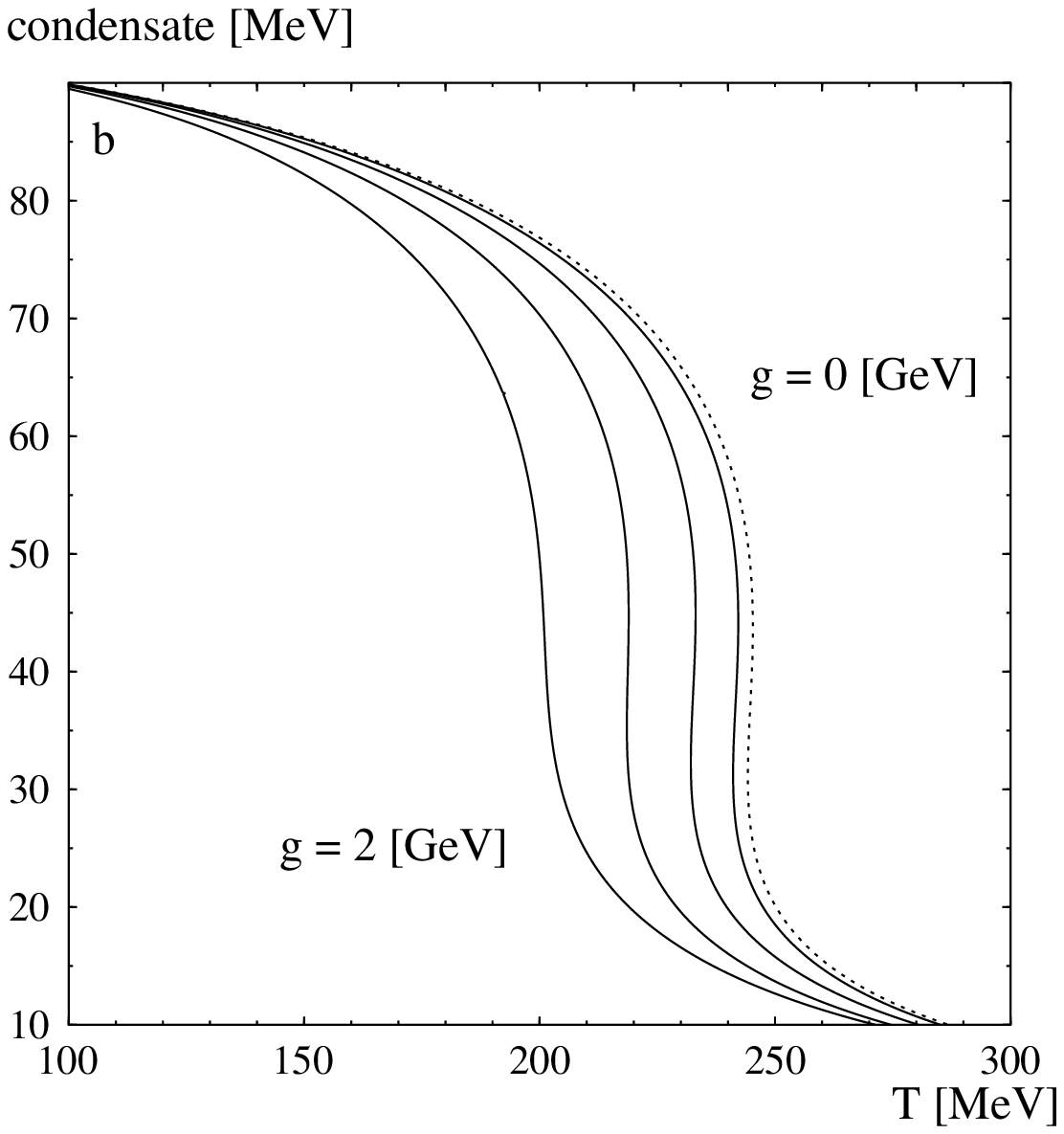}
\end{center}
\caption{ Panel (a): Order of the phase transition as a function of the
parameters of the model. $M_{H}=0.4$ GeV and $M_{S}$ and $g$ are varied. The
forbidden area violates the constraint $\left\vert
M_{S}^{2}-M_{H}^{2}\right\vert \geq 4g\protect\varphi _{0}$ \protect\cite%
{tqft}. On the border line between the first-order and the crossover
transitions a second-order phase transition is realized. Panel (b): the
chiral condensate is shown for $M_{H}=0.4$ GeV and $M_{S}$=1.0 GeV for
different values of $g$ (step of $0.5\text{ GeV)}$. The dots in panel
(a) correspond to the curves in panel (b).}
\label{22}
\end{figure}

We now turn to the explicit evaluation of masses, condensates, and the
mixing angle at nonzero $T$. The masses are chosen to be in the range quoted
by Refs.\ \cite{pdg,pelaez}: $M_{S}=1.2\text{ GeV}$ and $M_{H}=0.4\text{ GeV}$. The
coupling strength is set to $g=3.4\text{ GeV}$ in order to obtain a
crossover phase transition. Together with the pion mass $M_{\pi }=139\text{
MeV}$ and the pion decay constant $\varphi _{0}=f_{\pi }=92.4\text{ MeV}$
the parameters are determined as: $\lambda =52.85$, $M_{\chi }=0.96\text{ GeV}
$, and $F=64.2\text{ MeV}$.

\begin{figure}[tbp]
\begin{center}
\includegraphics[scale=0.35] {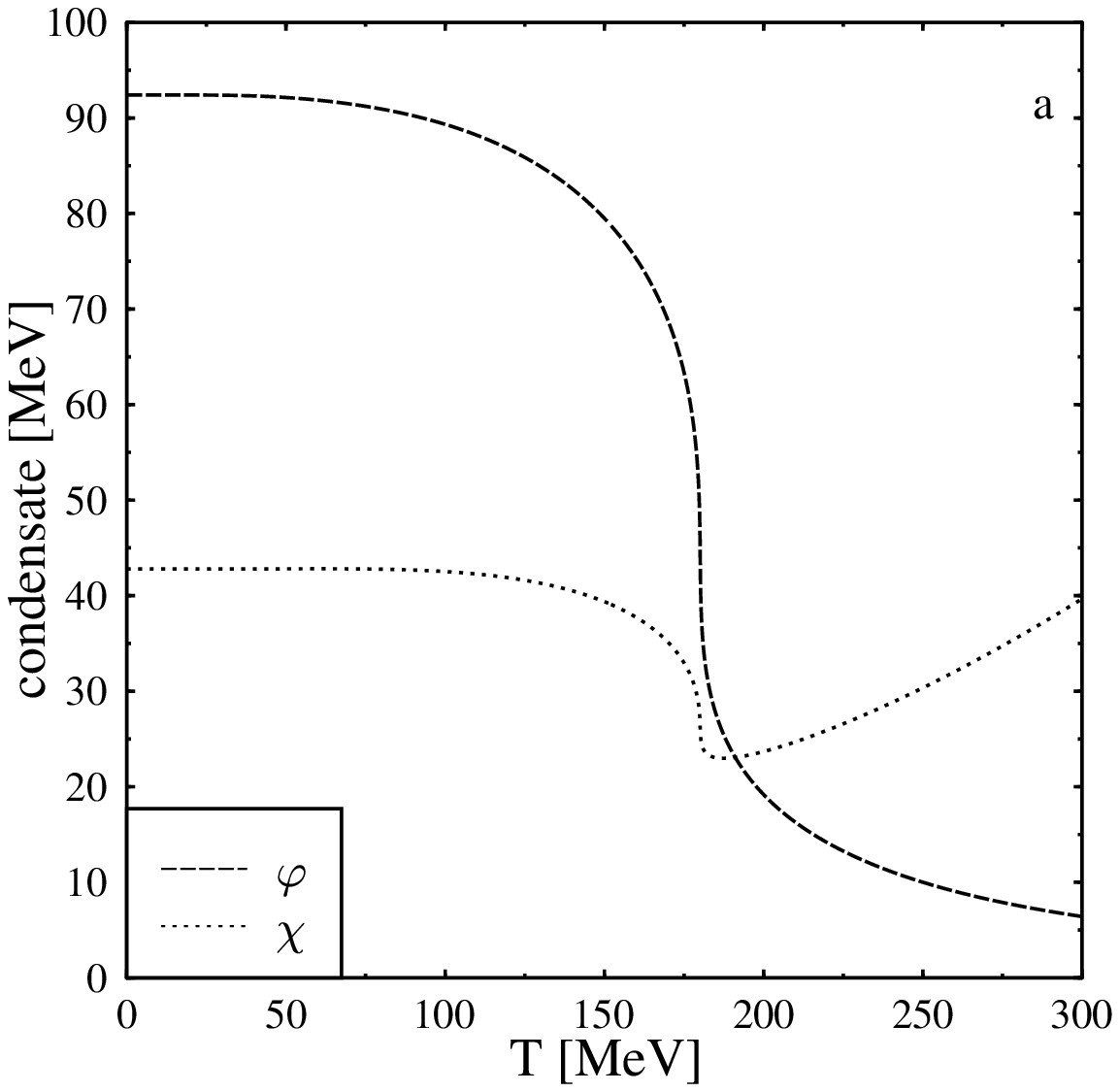}%
\includegraphics[scale=0.35]
{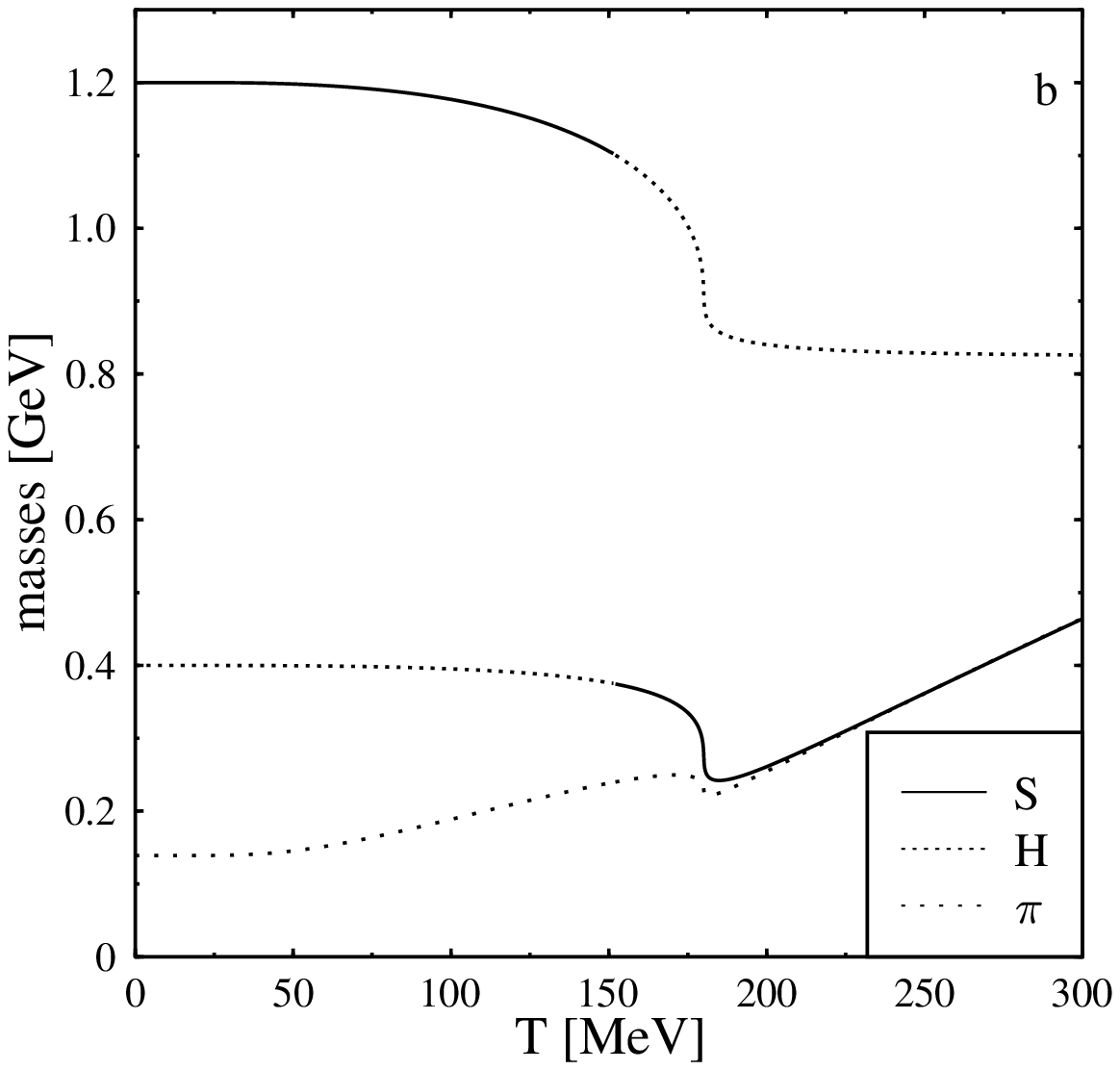}\includegraphics[scale=0.35] {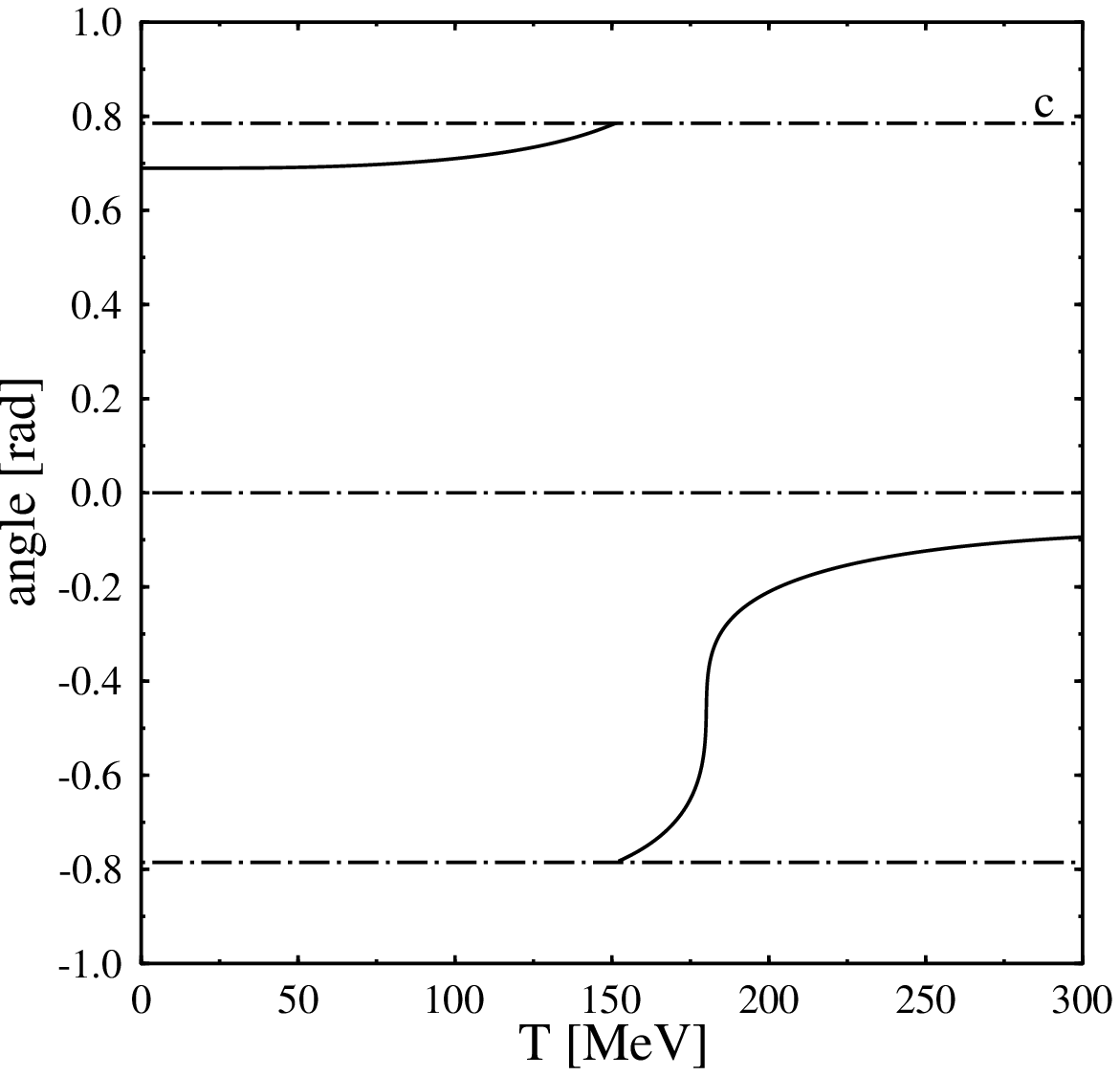}
\end{center}
\caption{Condensates (panel a), masses (panel b), and mixing angle (panel c)
as function of $T$. (From Ref.\ \protect\cite{tqft}.)}
\end{figure}

The behavior of the two condensates is shown in Fig.\ 2.a. At $T_{c}=180\text{
MeV}$ the quark condensate $\varphi (T)$ drops and approaches zero, thus
restoring chiral symmetry. Below $T_{c}$ the tetraquark condensate $\chi (T)$
follows the quark condensate, but above $T_{c}$ the condensate starts to
increase. (This result could be different if additional terms $\sim \chi ^{4}
$ in Eq.\ (\ref{pot}) were included.)

By increasing $T$ the function $M_{S}(T)$ first drops softly, but at a
certain temperature $T_{s}\simeq 160\text{ MeV}$ a step-like decrease
occurs, while the function $M_{H}(T)$ undergoes a step-like increase. The
solid line in Fig.\ 2.b describes the state $S$ according to the following
criterion: $S$ is the state containing the largest amount of the bare
quarkonium state $\varphi $. For $T<T_{s}$ it corresponds to the heavier
state, for $T>T_{s}$ to the lighter one. A similar analysis holds for the
dashed line referring to $H$ as the state with the largest bare tetraquark
amount.

The mixing angle $\theta (T)$ is shown in Fig.\ 3.c. At $T_{s}$ the mixing
becomes maximal and the angle jumps suddenly from $\pi /4$ to $-\pi /4$, $%
\lim_{T\rightarrow T_{s}}=\mp \pi /4$. At $T_{s}$ the two physical states $H$
and $S$ have the same amount (50\%) of quarkonium and tetraquark. 

\section{Conclusions}

\label{4}

We have shown that the interpretation of $f_{0}(600)$ as a predominantly
tetraquark state sizably affects the thermodynamical properties of the chiral
phase transition: the behavior of the quark condensate is softened
rendering the order of the phase transition cross-over for a sufficiently large 
tetraquark-quarkonium interaction,
and the value of the critical temperature $%
T_{c}$ is reduced, in agreement with recent Lattice simulations \cite%
{karschfodor}.

In future studies one should include a complete treatment of the other
scalar-isoscalar states $f_{0}(980)$, $f_{0}(1500)$, and $f_{0}(1710)$ which
appear in an $N_{f}=3$ context (together with the inclusion of the scalar
glueball). Also, (axial-)vector mesons shall be considered \cite{kr}.
Nevertheless, the emergence of mixing of tetraquark and quarkonium states is
general, and it is expected to play a relevant role at nonzero temperature also in
this generalized context.

\end{document}